\documentclass[]{article} 

\usepackage[utf8]{inputenc}
\usepackage{graphicx}
\usepackage{eurosym}

\usepackage[nonatbib, final]{neurips_2019}
\usepackage{tikz}

\usepackage[T1]{fontenc}    
\usepackage{hyperref}       
\usepackage{url}            
\usepackage{booktabs}       
\usepackage{amsfonts}       
\usepackage{nicefrac}       
\usepackage{microtype}      

\usepackage{amsthm, amsmath, amssymb, amsfonts}
\usepackage{enumerate}
\usepackage{mathrsfs}
\usepackage{microtype}
\usepackage{aliascnt}
\usepackage{graphicx}
\usepackage{mathtools}
\usepackage{caption}
\usepackage{wrapfig}

\usepackage{boldline,multirow}

\usepackage{footnote}
\makesavenoteenv{tabular}
\makesavenoteenv{table}

\usepackage{shorthands}

\title{Deep Hedging: \\Learning to Simulate Equity Option Markets}

\author{%
  Magnus Wiese
  \\
  \And Lianjun Bai
  \\
  \And Ben Wood\thanks{Corresponding author: \texttt{ben.wood@jpmchase.com}}
  \\ 
  \vspace{0.5cm}
  \\ \hspace{-3.2cm} J.P. Morgan\thanks{Opinions expressed in this paper are those of the authors, and do not necessarily reflect the view of J.P. Morgan.}
  \\
  \And Hans Buehler
}

\date{\today}

\begin{document}
\maketitle
\begin{abstract}
We construct realistic equity option market simulators based on generative 
adversarial networks (GANs). We consider recurrent and temporal convolutional
architectures, and assess the impact of state compression.
Option market simulators are highly relevant because they allow us 
to extend the limited real-world data sets available for the training and evaluation of option 
trading strategies. 
We show that network-based generators outperform classical methods on a range of 
benchmark metrics, and adversarial training achieves the best performance. Our work demonstrates
for the first time that GANs can be successfully applied to the task of generating multivariate financial time series.
\end{abstract}

\section{Introduction}
There is growing interest in applying reinforcement learning techniques to the problem
of managing a portfolio of derivatives \cite{Buehler2019, Ritter2018}. This involves
buying and selling not only the relevant underlying assets, but also the available
exchange-traded options. In order to train an option trading model, we therefore require 
time-series data that includes option prices. 

Unfortunately, the amount of useful real-life data available is limited; if we
take a sampling interval of one day, ten years of option prices translates into 
only a few thousand samples. This motivates the need for a realistic simulator: we can 
generate much larger volumes of data for training and evaluation while preserving
the key distributional properties of the real samples, thus helping to avoid overfitting.

In this article, we build and test a collection of simulators based on neural networks (NNs).
We begin by transforming the option prices to an equivalent representation in the form of \emph{discrete local volatilities} (DLVs) \cite{Buehler2017, Wissel2007} with less 
complicated no-arbitrage constraints; we then formalize the role of the simulator as a 
mapping from a state and input noise to a new set of (transformed) option prices.
Next, we define a series of benchmark scores based on the key distributional features 
of the transformed prices. We construct a set of generative models, varying the 
network architecture, training method, and state compression scheme. 
Finally, we evaluate these models against our proposed benchmark scores, and 
compare with a classical baseline.

\section{Financial time series simulation}
\label{sec:related_work}
There is a wide range of existing literature on the generation of synthetic time
series for asset prices (see, e.g., \cite{Francq2010}). Classical derivative pricing models 
(e.g., \cite{BlackScholes1973, Dupire1994, Heston93}) also require 
path generators, but these are not designed to be realistic; they describe diffusion in the 
risk-neutral measure $\mathbb{Q}$, rather than the real-world measure $\mathbb{P}$, and are
typically limited to a small number of driving factors, for ease of computation.

Recently, generative adversarial networks (GANs) \cite{Goodfellow2014} have been successfully
used to create realistic synthetic time series for asset prices 
\cite{Koshiyama2019, Takahashi2019, Wiese2019, Zhang2019, Zhou2018}. 
Zhang et al.~\cite{Zhang2019} and Zhou et al.~\cite{Zhou2018} reported on using the objective 
function of GANs to predict spot prices. Koshiyama et al.~\cite{Koshiyama2019} trained a conditional 
GAN to generate spot log-returns and provided a study of using generated paths to fine-tune trading 
strategies, and showed that the autocorrelation function (ACF) and partial autocorrelation function 
(PACF) could be generated accurately by using a two-layer perceptron. Takahashi et 
al.~\cite{Takahashi2019} also reported on being able generate various stylised facts \cite{Cont2002} 
found in the historical series, but did not provide a detailed description of the methodology used. 
An unconditional approach to the generation of spot price log-returns, using Temporal 
Convolutional Networks (TCNs) \cite{Oord2016}, was first presented by Wiese et al.~\cite{Wiese2019}; 
they also reproduce the relevant stylised facts, including volatility clustering and the leverage effect. 

In contrast, generative models for time series of option prices are much less common:
Cont \cite{Cont2002} performs a principal component analysis (PCA) on implied volatility data; 
Wissel \cite{Wissel2007} provides a scheme to build a risk-neutral market model, focusing on 
ensuring the martingale property rather than realism. As far as we are aware, neural networks have not 
previously been applied to option market generation. 

Our work extends the conditional modelling framework of \cite{Koshiyama2019} to the multivariate 
setting by using GANs and other calibration techniques. 

\section{Option prices}
Our aim in this article is to simulate the prices of standard ``vanilla'' equity index options, of the type commonly 
traded on exchanges in large volumes\footnote{Eurex \cite{Eurex2019} reports an average of more than one 
million option contracts traded daily on EURO STOXX 50 in August 2019, corresponding to almost 
\EUR{650MM} in premiums paid.}. An option is characterized by: the underlying index; the type (\emph{call} or 
\emph{put}); the {strike}, $K$; and the {maturity}, $T$. The maturity is the expiry date of the option; 
on that date, the option holder receives an amount $\max\bigl(0,s(I_T-K)\bigr)$, where $I_T$ is the prevailing 
level of the underlying index, and $s=1$ for a call and $-1$ for a put. 

At any given time, not all strike/maturity/type combinations are tradable; market makers quote bid and/or offer
prices for only the most relevant combinations, which broadly means those with strike closest to the current index 
level, and maturity closest to today. We will work with a representative grid of strikes and maturities. 

\begin{figure*}[htp]
        \centering
        \includegraphics[width=\textwidth]{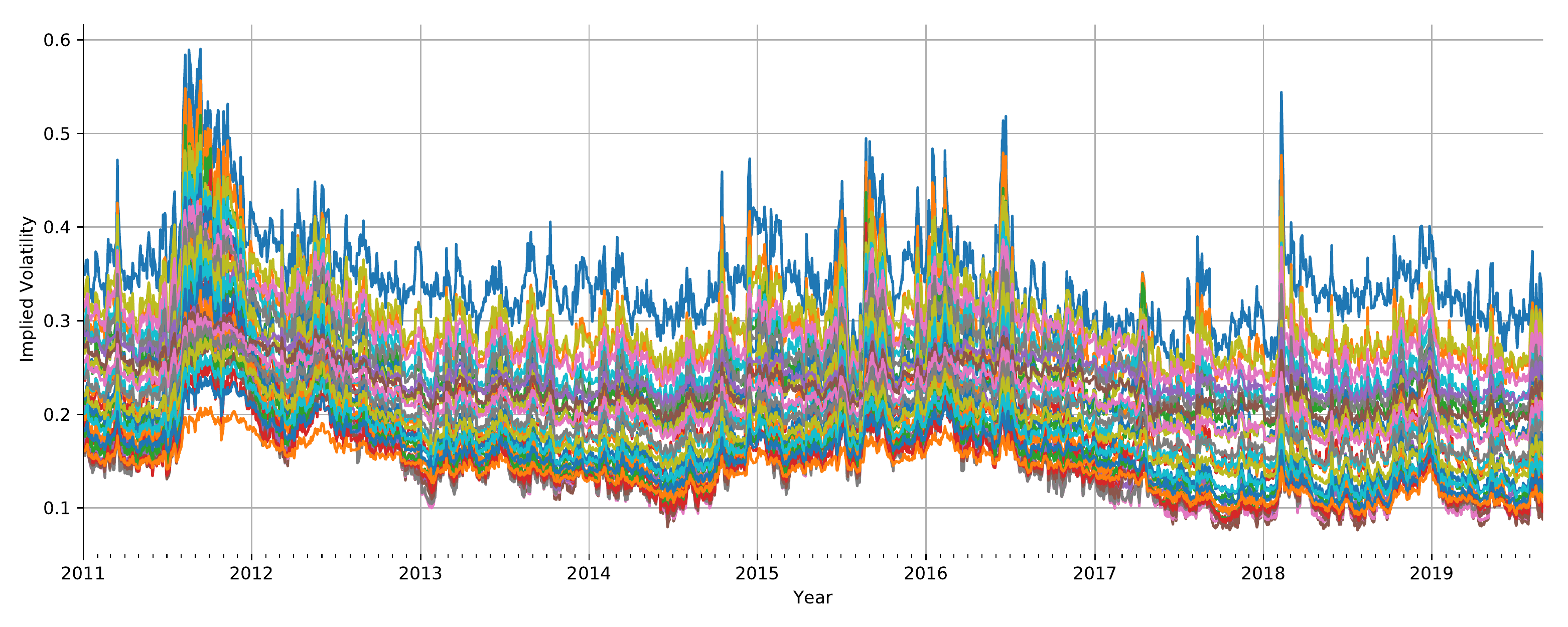}
        \caption{Implied volatilities of the EURO STOXX 50.}
        \label{fig:historical_implied_vols}
\end{figure*}

Market participants commonly express option prices in terms of \emph{implied volatilities}. The implied volatility
is the number one must use in the standard Black-Scholes formula \cite{BlackScholes1973} to obtain the option price. 
The other inputs to the formula are the discount factor and the index forward price. All three inputs to the price
are stochastic, but the short-term price dynamics are primarily driven by changes in the implied volatility; in this 
paper, we focus on this contribution.

Option prices are subject to strict ordering constraints because of no-arbitrage considerations. For example, 
since the call option payoff is a non-increasing function of strike, the option price must also be non-increasing;
violation of this rule would constitute an arbitrage opportunity\footnote{Strictly, there is an arbitrage opportunity
when the bid price of the higher-strike call option exceeds the offer price of the lower-strike option.}, i.e.
the possibility of making a profit while guaranteeing no loss. True arbitrage opportunities are rare, 
fleeting, and small; an option price simulator is more useful if it does not generate arbitrageable market
states.

For this reason, it is not convenient to work with option prices directly; the ordering constraints are too 
complicated. The equivalent constraints on implied volatilities are even more awkward.
Instead, we transform the prices to DLVs 
\cite{Buehler2017, Wissel2007}, for which the absence of arbitrage corresponds to a simple requirement of 
positivity. In the absence of arbitrage, this mapping is bijective; an $N_K \times N_M$ grid of option prices is 
converted into an $N_K \times N_M$ grid of DLVs.

In this article we will focus on one of the largest option markets, namely options on the EURO STOXX 50 index. An overview of the distributional and dependence properties of DLVs is provided in \autoref{appendix:DLVs}. The implied volatilities of the EURO STOXX 50 for the relevant sets of strikes and maturities are displayed in \autoref{fig:historical_implied_vols}. 

\section{Problem formulation}
\label{sec:setup}
Throughout this paper $\mathbb{N}_0$ is the time set, $(\Omega, (\mathcal{F}_t)_{t\in \timeset}, \mathbb{P})$ the filtered probability space and $L^2(\mathbb{R}^N)$ denotes the space of $\mathbb{R}^N$-valued measurable functions for which the Euclidean norm $\|\cdot\|_2$ is Lebesgue integrable. 

We assume a set of $N_K$ equispaced strikes 
\[
\strikes=\left\lbrace K_1, K_1 + \Delta K, \dots, K_1 + (N_K - 1) \Delta K \right \rbrace
\]
and a set of $N_M$ maturities 
\[
\maturities = \left\lbrace M_1, \dots, M_{N_M}\right\rbrace
\]
for which we obtain the $(N_K \cdot N_M)$-dimensional process of DLVs 
\begin{equation*}
\dlv_t \coloneqq [\dlv_t(K, M)]_{(K, M) \in \strikes\times\maturities}, \ \textrm{for} \ t \in \timeset.
\end{equation*}
Furthermore, we assume that the historical process $\processhist$ evolves through a conditional model which can be constructed by feeding in a state $S_t$, which we would like to condition on, and noise $Z_{t+1}$ which drives the process. Particularly, we describe the evolution of $\processhist$ by a unkown mapping $g:L^2(\mathbb{R}^{N_Z})\times L^2(\mathbb{R}^{N_S}) \to L^2(\mathbb{R}^{N_K \cdot N_M})$ which relates noise and state to the next time step, such that our process takes the form
\begin{align}
\label{eq:objective}
\dlv_{t+1} = g(Z_{t+1}, S_t), \ \textrm{for} \ t \in \timeset
\end{align}
where $Z_{t+1}\sim \mathcal{N}(0, I)$ is $N_Z$-dimensional Gaussian noise and the state $S_t$ a function of the processes history, i.e. $S_t = f(\dlv_t, \dots, \dlv_0)$. An example of $f$ is a projection onto the most current component, i.e. $S_t = \dlv_t$, the last $L$ realisations, $S_t = [\dlv_t, \dots, \dlv_{t-L}]$ or an exponentially weighted moving average.

The objective is to approximate the mapping $(Z_{t+1}, S_t) \mapsto \dlv_{t+1}$ which ideally allows us to generate more data from a given state $S_t$. In this paper our approach is to represent this mapping through a deep neural network
\[ 
g: L^2\left(\mathbb{R}^{N_Z}\right) \times  L^2\left(\mathbb{R}^{N_S}\right) \times \Theta \to L^2\left(\mathbb{R}^{N_K \cdot N_M}\right)
\]
which is defined as a function of noise, state and its parameters $\theta \in \Theta$. For a given parameter vector $\theta \in \Theta$ the generated process $\processgen$ is defined as
\begin{equation}
\label{eq:dnn_model}
\dlvgen_{t+1, \theta} = g_\theta(Z_{t+1}, \tilde S_{t, \theta})
\end{equation}
where $\tilde S_{t, \theta} = f(\dlvgen_{t, \theta}, \dots, \dlvgen_{0, \theta})$. The optimal outcome is to approximate a parameter vector $\theta_{\textrm{ML}} \in \Theta$ such that $(\dlv_t, t \in \timeset)$ and $(\dlvgen_{t, \theta_{\textrm{ML}}}, t \in \timeset)$ inherit the same dynamics in terms of distributional and dependence properties.

\section{Models}
\label{sec:generative_models}
We now turn to the problem of modeling the process of DLV levels $\processhist$ and consider from now on log-DLV levels $\logdlv_t \coloneqq \log(\dlv_t) $ for $t \in \timeset$ to ensure non-negativity of the generated time series. 

The first (naive) approach to model log-DLVs is to approximate \eqref{eq:dnn_model} directly and generate all DLVs yielding the generated time-series 
\begin{equation}
\label{eq:explicit_model}
\logdlvgen_{t+1, \theta} = g_\theta(Z_{t+1}, \tilde S_{t, \theta})
\end{equation}
where $\tilde S_{t, \theta} = [\logdlvgen_{t, \theta}, \dots, \logdlvgen_{t-L, \theta}]$ includes current and past log-DLV levels for tuneable $L \in \mathbb{N}$. However, this approach suffers from having to model an arbitrarily high-dimensional process when it is necessary to generate a fine grid of DLVs yielding $N_X \gg 1$ for $N_X\coloneqq N_K \cdot N_M$. Consequently, we also explore a compressed version of our generator. 

\begin{figure}[htp]
    \centering
    \begin{minipage}{.45\textwidth}
       \centering
	\includegraphics[width=\textwidth]{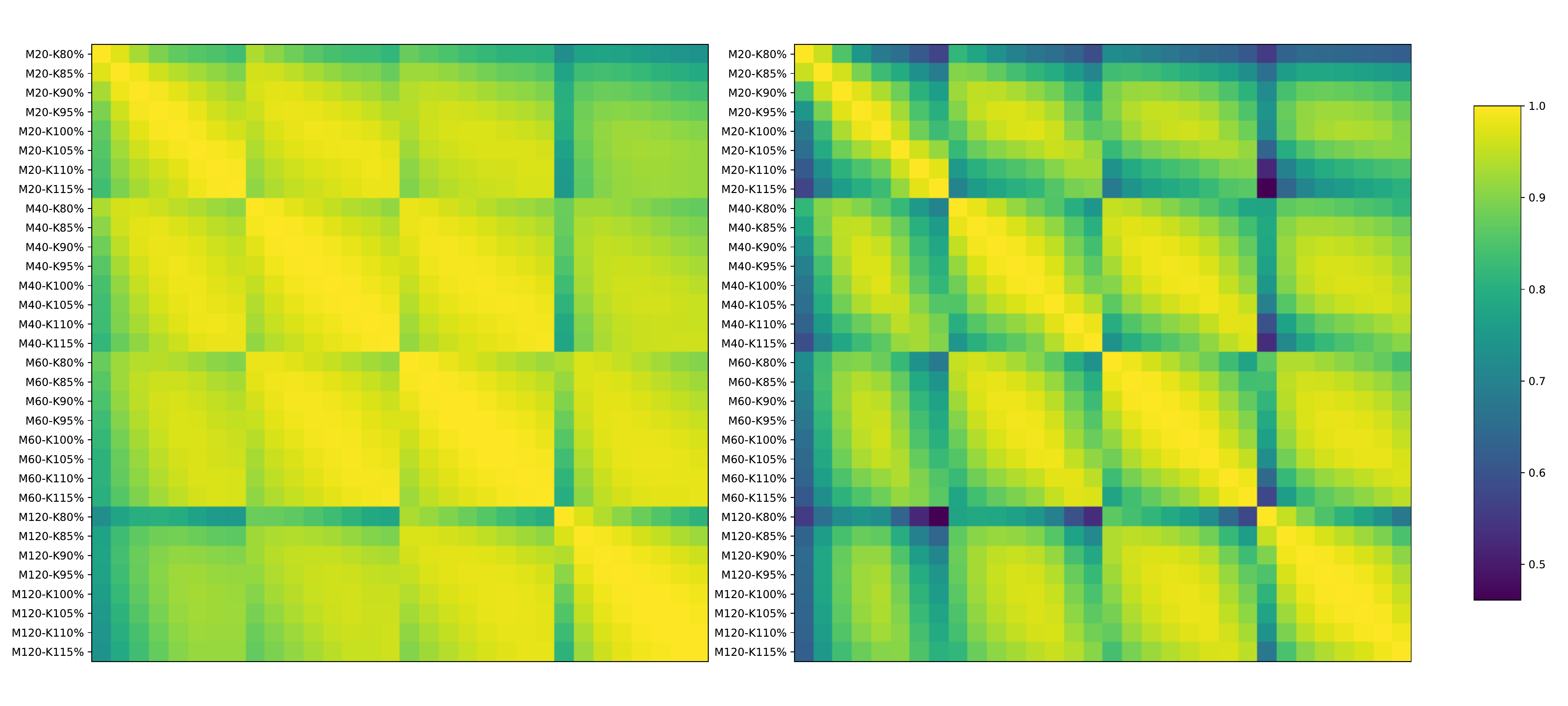}
	\caption{Cross-correlation matrix of log-DLV levels (left) and DLV log-returns (right). Labels on the $y$-axis indicate the maturity ($M$) and relative strike ($K$) of each row.}
	\label{fig:cross_corr}
    \end{minipage}
	\hfill
    \begin{minipage}{0.45\textwidth}
       \centering
	\includegraphics[width=\textwidth]{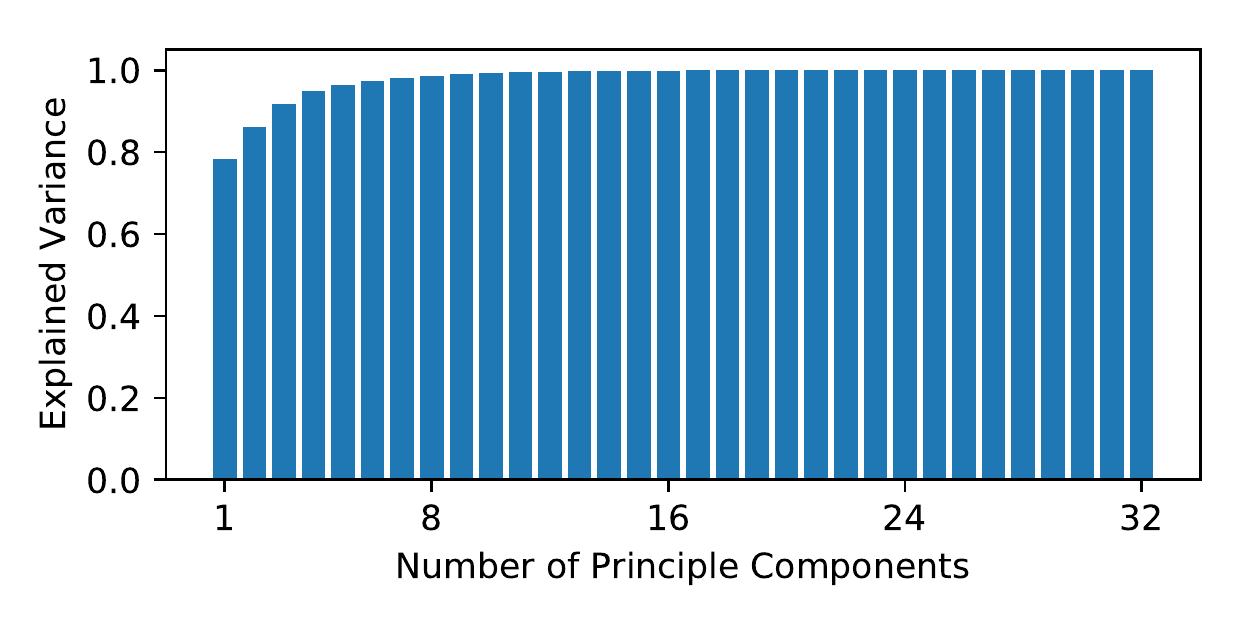}
	\caption{Cumulative sum of variance explained by the first 10 principal components.}
	\label{fig:explained_variance}
    \end{minipage}
\end{figure}

The compressed version is motivated from the observation that log-DLV levels have high cross-correlations indicating that log-DLVs live on a lower-dimensional manifold. \autoref{fig:cross_corr} illustrates the cross-correlation matrix of log-DLVs and DLV log-returns for the set of relative strikes ${\bar\strikes \coloneqq \lbrace 0.80, 0.85, 0.90, 0.95, 1.00, 1.05, 1.10, 1.15\rbrace}$ and maturities $\bar\maturities \coloneqq \lbrace 20, 40, 60, 120 \rbrace$. Both cross-correlation matrices show high cross-correlations between groups of different maturities as well as within a specific maturity. 

We therefore apply PCA to the set of log-DLV levels $(\logdlv_t, t\in\timeset)$ and compress the $N_X$-dimensional process into an $N_\pca$-dimensional one. The compressed process is defined for $t \in \timeset$ as $\pca_t = \mathbf{W} \logdlv_t$, where $\mathbf{W} \in \mathbb{R}^{N_\pca \times N_X}$ denotes the compression matrix obtained through PCA. The generated process now takes the form
\begin{align*}
\tilde \pca_{t+1, \theta} &= g_\theta(Z_{t+1}, \tilde S_{t, \theta})
\\
\logdlvgen_{t+1, \theta} &= \mathbf{W}_{}^\dagger\tilde \pca_{t+1, \theta}
\end{align*}
where $\tilde S_{t, \theta} = [\tilde \pca_{t, \theta}, \dots, \tilde \pca_{t-L, \theta}]$ is the state of current and past compressed log-DLVs - again for tuneable $L\in \mathbb{N}$.

We also explored compressing DLV log-returns $R_t \coloneqq \logdlv_t - \logdlv_{t-1}, \ t \in \timeset$ as they are also highly correlated (see \autoref{fig:cross_corr}). However, we discarded this approach since it involves taking the cumulative sum of the compressed DLV log-returns which causes a lossy compression for a small number of principle components. 

\autoref{fig:explained_variance} illustrates the cumulative sum of variance explained as a function of the number of principle components for the grid $\bar\strikes \times \bar\maturities$. In \autoref{sec:numerical_results} we report our results for $N_P = 5$ principle components which explain $\approx 96\%$ of the variance yielding a good trade-off between dimensionality reduction and preserving information.


\section{Optimization}
The next step is to obtain a close approximation of $\theta_{\textrm{ML}}$. Various training procedures exist ranging from conventional parametric methods such as quasi maximum likelihood estimation (qMLE) to novel non-parametric ones such as GANs and the mean maximum discrepancy (MMD) \cite{Sutherland2017}. In this paper, we explore the performance of qMLE, GANs and Wasserstein GANs (WGAN-GP) \cite{Gulrajani2017}. Following, we provide a brief explaination of the qMLE and GANs in the context of our explicit conditional model \eqref{eq:explicit_model}.

In qMLE we assume that the distribution of $  X_{t+1} |  (Z_{t+1}, S_{t}) $ can be described by a family of distributions $\mathcal{P} = \lbrace \mathbb{P}_\gamma \ | \ \gamma \in \Gamma \rbrace $ for some parameter space $\Gamma$. The objective is to maximize the likelihood of our generated data under our family $\mathcal{P}$ \cite[Chapter 5]{Goodfellow2016}. 
Here, we assume that $  X_{t+1} |  (Z_{t+1}, S_{t}) $ follows a Gaussian distribution where we constrain our covariance matrix to be diagonal. Our parametric family thus takes the form
\begin{equation*}
\mathcal{P} =\lbrace \mathcal{N}(\mu, \Sigma) \ | \ \mu \in \mathbb{R}^{N_X},  \Sigma \in \mathbb{D} \rbrace.
\end{equation*}
where $\mathbb{D}\coloneqq\lbrace\operatorname{diag}(a_1, \dots, a_{N_X}) \ | \ a_1, \dots, a_{N_X} \in \mathbb{R}_{\geq 0}\rbrace$ is the set of $(N_X \times N_X)$-dimensional diagional matrices with non-negative components. 

A challenge in qMLE is the correct specification of $\mathcal{P}$ and the intractability of likelihood functions. GANs try to address this issue by introducing a min-max two-player game between the generator $g_\theta$ and the discriminator $d_\eta$. The discriminator aims to discriminate between real samples from the data distribution and synthetic ones generated by the generator. The objective function proposed by the original paper from Goodfellow \cite{Goodfellow2014} adapted to our setting is of the form 
\begin{equation*}
	\mathcal{L}(\theta, \eta) = \expectation\left(\log(d_\eta([S_t, X_{t+1}]))\right) + \expectation\left(\log(1-d_\eta([\tilde S_{t, \theta}, \tilde X_{t+1,\theta}]))\right)
\end{equation*}
where $\tilde X_{t+1,\theta}$ is defined as in \eqref{eq:explicit_model}. 

A drawback of GANs is that they are notoriously hard to train which lead to the introduction of various regularization techniques to stabilize training \cite{Arjovsky2017, Brock2018, Gulrajani2017, Mescheder2018, Miyato2018, Sonderby2016}. In our numerical results, we grid search over spectral normalization \cite{Miyato2018} imposed on the discriminator and generator \cite{Brock2018} and gradient penalities proposed by Mescheder \cite{Mescheder2018}.
Furthermore, we also train our generative model by using WGAN-GP proposed by ~\cite{Gulrajani2017} and report on these results separately in \autoref{sec:numerical_results}. 

\section{Evaluating the generated paths}
\label{sec:evaluation}
In GANs the objective function cannot be used to evaluate the performance of the generator. Equally, using the likelihood of a qMLE-trained model can give a distorted image. To measure the goodness and performance of a generative model we define and introduce various metrics and scores. These scores allow us to capture whether the generator is able to generate dynamics that are similar to those found in the historical DLV series such as highly cross-correlated log-DLVs and DLV log-returns, bimodal distributions or persistence in the autocorrelation. 

During training we intentionally evaluate log-DLVs instead of implied volatilities due two reasons. First, a close approximation of the generating mechanism of DLVs yields a close approximation of the generating mechanism of implied volatilities. Second, transforming DLVs to implied volatilities is costly. Once a close approximation of $\theta_{\textrm{ML}}$ is obtained we transform the generated DLVs to implied volatilities and compute metrics and scores for the generated implied volatilities and report on those. 

We denote the historical dataset of log-DLVs by $\mathcal{D}_h= \left\lbrace [x_0, \dots, x_T] \right\rbrace$ and likewise the generated dataset containing $M \in \mathbb{N}$ paths of length $T$ through
\[
\mathcal{D}_g = \left\lbrace[\tilde x^{(i)}_{0, \theta}, \dots, \tilde  x^{(i)}_{T, \theta}]\right\rbrace_{i=1}^{M}
\]
where $[\tilde x^{(i)}_{0, \theta}, \dots, \tilde  x^{(i)}_{T, \theta}]$ denotes for any $i \in\lbrace1, \dots, M\rbrace$ a time series obtained through recursive sampling from an initial state sampled from the historical dataset $\mathcal{D}_h$.

We begin by introducing a distributional metric and distributional scores, then define dependence scores and at last two scores that take into account the cross-correlation structure. 

\subsection{Distributional metric}
Naturally, we want the unconditional distribution of the generated and historical to match closely. For this purpose, let $\mathcal{B}_h = \left\lbrace B_1, \dots, B_{K}\right\rbrace$ be a binning such that approximately $20$ elements of the historical series $x \in \mathcal{D}_h$ fall into each bin; $\# \left\lbrace t \in \left\lbrace 0, \dots, T \right\rbrace: x_{t}  \in B \right\rbrace \approx 20$ for any $ B \in \mathcal{B}_h$. With respect to the binning the empirical probability density function (epdf) of the historical ${\hat f_h: \mathcal{B}_h \to \mathbb{R}_{\geq 0}}$ and the generated $\hat f_g: \mathcal{B}_h \to \mathbb{R}_{\geq 0}$ can be defined. 
During training we track the absolute difference of the epdf
\[
\sum_{B \in \mathcal{B}_h} |\hat f_h(B) - \hat f_g(B)|.
\]
\subsection{Distributional scores}
In financial applications higher order moments such as the skewness and kurtosis are of interest as they determine the propensity to generate extremal values. We therefore define the skewness score
\[
\frac{1}{N_X}\sum_{j=1}^{N_X}\left\| \skewness({x_{:, j}})  - \skewness\left({[\tilde x_{:,\theta, j}^{(1)}, \dots, \tilde x_{:, \theta, j}^{(M)}]}\right)\right\|_2
\]
where $\tilde x_{:,\theta, j}^{(i)}$ for $i \in \{1, \dots, M\}$ denotes the $j$-th dimension of the $i$-th generated time series and likewise the kurtosis score
\[
\frac{1}{N_X}\sum_{j=1}^{N_X}\left\| \kurtosis({x_{:, j}})  - \kurtosis\left({[\tilde x_{:,\theta, j}^{(1)}, \dots, \tilde x_{:, \theta, j}^{(M)}]}\right)\right\|_2.
\]

\subsection{Dependence scores}
Since DLVs are persistent we adopted ACF score proposed in \cite{Wiese2019}. It is defined by taking the Euclidean norm of the difference of the historical and the mean generated autocorrelation function
\[
\operatorname{ACF}_X \coloneqq 
\|
\operatorname{ACF}(x) - \frac{1}{|\mathcal{D}_g|}\sum_{\tilde x_{:, \theta} \in \mathcal{D}_g} \operatorname{ACF}(\tilde x_{:, \theta})
\|_2.
\]
Similarly, we define the ACF score for the log-return process $r_{t} = x_{t} - x_{t-1}, \ t \in \lbrace 1, \dots, T \rbrace$ 
\[
\operatorname{ACF}_R \coloneqq
\|
\operatorname{ACF}(r) - \frac{1}{|\mathcal{D}_g|}\sum_{\tilde r_{:, \theta} \in \mathcal{D}_g} \operatorname{ACF}(\tilde r_{:, \theta})
\|_2.
\] 
In \autoref{sec:numerical_results} we report the $\operatorname{ACF}_X$ score for the first $32$ lags and the $\operatorname{ACF}_R$ score for the first lag. 
\subsection{Cross-correlation scores}
To capture whether the generator generates cross-correlated log-DLVs and DLV log-returns we introduce two more scores. The cross-correlation score of log-DLVs is defined by taking the Euclidean norm of the cross-correlation matrix of log-DLVs ${\| \hat \Sigma_h^X - \hat \Sigma_g^X \|_2}$ where $\hat \Sigma^X_h, \hat \Sigma^X_g $ denote the cross-correlation matrix of the historical and generated respectively. Likewise, the cross-correlation score of DLV log-returns is defined ${\| \hat \Sigma_h^R - \hat \Sigma_g^R \|_2}$ where $\hat \Sigma^R_h, \hat \Sigma^R_g $ denote the cross-correlation matrix of the historical and generated DLV log-returns respectively.

\section{Numerical results}
\label{sec:numerical_results}
In this section we evaluate the performance of qMLE-, GAN- and WGAN-GP-trained models for the compressed and explicit version.

\subsection{Dataset}
We use call option prices of the EURO STOXX 50 from 2011-01-03 to 2019-08-30 and consider the set of relative strikes and maturities
\begin{align*}
\bar\strikes &\coloneqq \left\lbrace 0.80, 0.85, 0.90, 0.95, 1.00, 1.05, 1.10, 1.15\right\rbrace,\\
\bar\maturities &\coloneqq \left\lbrace 20, 40, 60, 120 \right\rbrace.
\end{align*}
For those option prices we compute the path of DLVs which we use for training (see \autoref{fig:historical_DLVs}). The total length of the time series is $\bar T\coloneqq 2257$. 
\subsection{Benchmarks}
For comparison we apply the vector autoregressive model $\var$ \cite{Johansen1995} and GAN-trained TCNs \cite{Wiese2019} to the same data. $\var$ is a standard model for multivariate time series and assumes that $X_{t+1}$ is an affine function of the past $p$ observations and some Gaussian noise $Z_{t+1}\sim\mathcal{N}(0, \Sigma)$:
\[
\logdlv_{t+1} = \mathbf{A}_1 \logdlv_{t} + \dots + \mathbf{A}_p \logdlv_{t-p} + \mathbf{b} + Z_{t+1}.
\]
TCNs model log-DLVs unconditionally. Here, the generated process takes the form
\[
\logdlvgen_{t+1, \theta} = g_\theta(Z_{t+1}, \dots, Z_{t+1-L})
\]
where $(Z_t, t\in\mathbb{Z})$ is an i.i.d. Gaussian process and $L\in\mathbb{N}_0$ denotes the TCN's receptive field size (see \cite{Wiese2019}). 

\subsection{Training and evaluation time}
We split the the historical series into a training and validation set through random sampling. The training set holds 85\% of the data and is used to calibrate the parameters of the model. During training we compute the scores for $\bar M\coloneqq40$ generated paths of length $\bar T$. GAN- and WGAN-GP-trained models are evaluated every 100 generator gradient updates. qMLE-trained models are trained through early stopping and are only evaluated once after the criterion has been reached. The VAR model is also only evaluated once after the parameters are obtained through regression. 

\subsection{Results}
\autoref{table:full_path} summarises the best scores obtained for each of the generative models for a fixed parameter vector. For all except the $\operatorname{ACF}_X$ score the explicit GAN-trained model achieves to generate the best paths. 

\begin{table}[htp]
\centering
\captionof{table}{Scores obtained from historical implied volatilities and generated paths.}
\resizebox{\linewidth}{!}{
\begin{tabular}{lccccccc}
\toprule
& \multicolumn{3}{c}{Distributional} & \multicolumn{2}{c}{Dependence} &  \multicolumn{2}{c}{Cross-Correlation} \\
\cmidrule(lr){2-4}\cmidrule(lr){5-6}\cmidrule(lr){7-8}
Models  & \multicolumn{1}{l}{$|\hat f_h(B) - \hat f_r(B)|$} & \multicolumn{1}{l}{skew} & \multicolumn{1}{l}{kurtosis} & \multicolumn{1}{l}{$\operatorname{ACF}_X$} & \multicolumn{1}{l}{$\operatorname{ACF}_R$} & \multicolumn{1}{l}{$\| \hat \Sigma^X_h - \hat \Sigma^X_g \|_2$} & \multicolumn{1}{l}{$\| \hat \Sigma^R_h - \hat \Sigma^R_g \|_2$} \\
\bottomrule
GAN             & \textbf{0.044} & \textbf{0.063} & \textbf{0.097} & 0.186 & \textbf{0.011} & \textbf{0.137} & \textbf{0.771}  \\
GAN - PCA(5)    & 0.047 & 0.082 & 0.276 & 0.468 & 0.021 & 0.977 & 3.125  \\
WGAN-GP           & 0.046 & 0.296 & 1.023 & 0.296 & 0.027 & 0.457 & 0.843  \\
WGAN-GP - PCA(5)   & 0.055 & 0.115 & 0.139 & \textbf{0.172} & 0.021 & 0.230 & 1.747  \\
qMLE            & 0.124 & 0.636 & 0.585 & 0.765 & 0.073 & 2.264 & 11.567 \\
qMLE - PCA(5)   & 0.075 & 0.427 & 0.463 & 1.048 & 0.020 & 2.434 & 3.599  \\
VAR(2)          & 0.088 & 0.415 & 0.447 & 0.834 & 0.013 & 1.113 & 2.770 \\
VAR(2) - PCA(5) & 0.223 & 0.328 & 2.302 & 0.914 & 0.025 & 0.477 & 1.982  \\
TCN(256)\footnote{The number in brackets specifies the receptive field size that was used; here 256.}        & 0.048 & 0.105 & 0.295 & 0.330 & 0.372 & 0.335 & 1.501  \\
\bottomrule
\end{tabular}}
\label{table:full_path}
\end{table}

The model that performs worst in terms of distributional properties is the VAR(2) - PCA(5) model. There the the fit of density and kurtosis is widely off. Notably, the $\operatorname{ACF}_X$ scores fit least for the VAR(2) and qMLE-trained models, independent whether they were trained on all or the compressed log-DLVs. Overall, we can also conclude from \autoref{table:full_path} that GAN- and WGAN-GP-calibrated models give the best fit. For TCNs we do not report on a compressed version as no good approximation could be obtained.

Although GAN-trained TCNs give a fairly good approximation in terms of distributional and cross-correlation scores the $\operatorname{ACF}_R$ score is far off. This makes the generated paths look very noisy. From this observation we concluded that TCNs have difficulties generating time series with high persistence from a pure i.i.d. noise process.

\subsection{Explicit GAN-trained model}
We presented numerical results for a wide range of models and optimization algorithms. Now, we will take a look at the properties of the explicit GAN-trained model since it performed best across most benchmark scores. 

\begin{figure}[!htb]
\begin{minipage}{.45\textwidth}
\centering
\includegraphics[width=\textwidth]{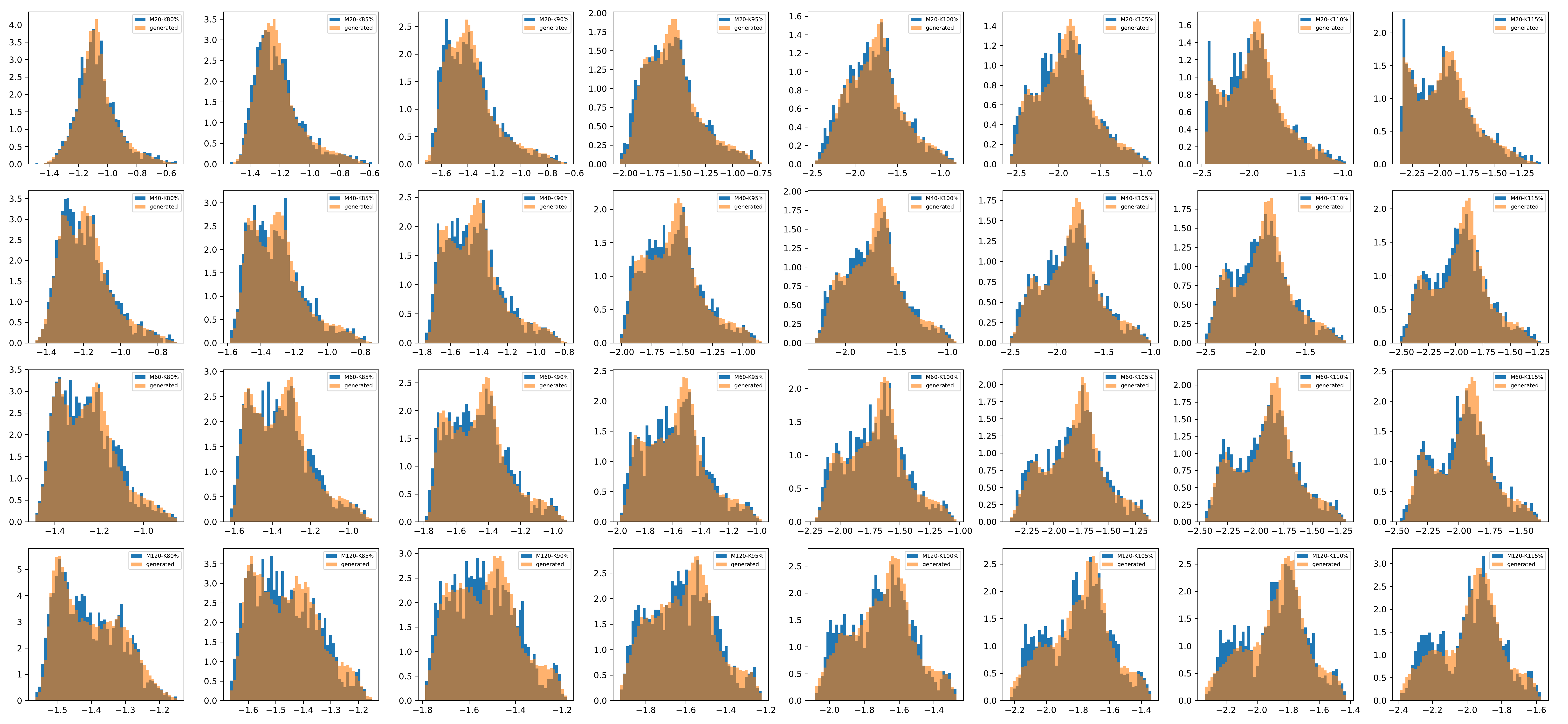}
\caption{Empircal densities of the historical (blue) and generated (orange) log-implied volatilities.}
\label{fig:explicit_gan_distributional}
\end{minipage}%
	\hfill
\begin{minipage}{.45\textwidth}
\centering
\includegraphics[width=\textwidth]{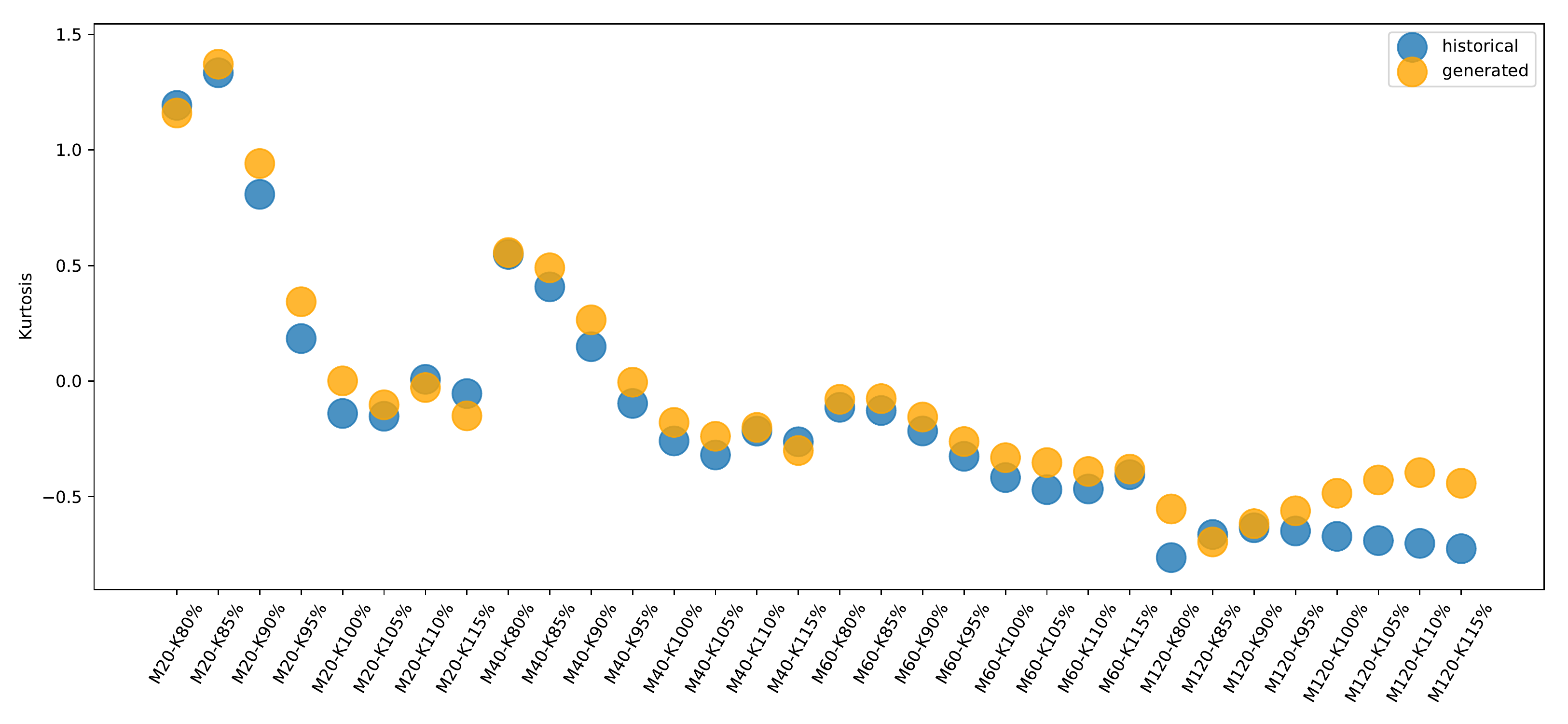}
\caption{Kurtosis of historical (blue) and generated (orange) log-implied volatilities.}
\label{fig:explicit_gan_kurtosis}
\end{minipage}
\end{figure}

As can be seen in \autoref{fig:explicit_gan_distributional} the epdfs of the historical and generated log-implied volatilities match closely for most maturities and relative strikes. Arguably, the fit of the bimodal distribution for long-dated ($M=120$ days) out of the money implied volatilies could be better. Taking a look at the historical and generated kurtosis in \autoref{fig:explicit_gan_kurtosis} we can conclude that for most implied volatilities the approximation is accurate. 

\begin{figure}[!htb]
    \centering
    \begin{minipage}{.45\textwidth}
        \centering
        \includegraphics[width=\textwidth]{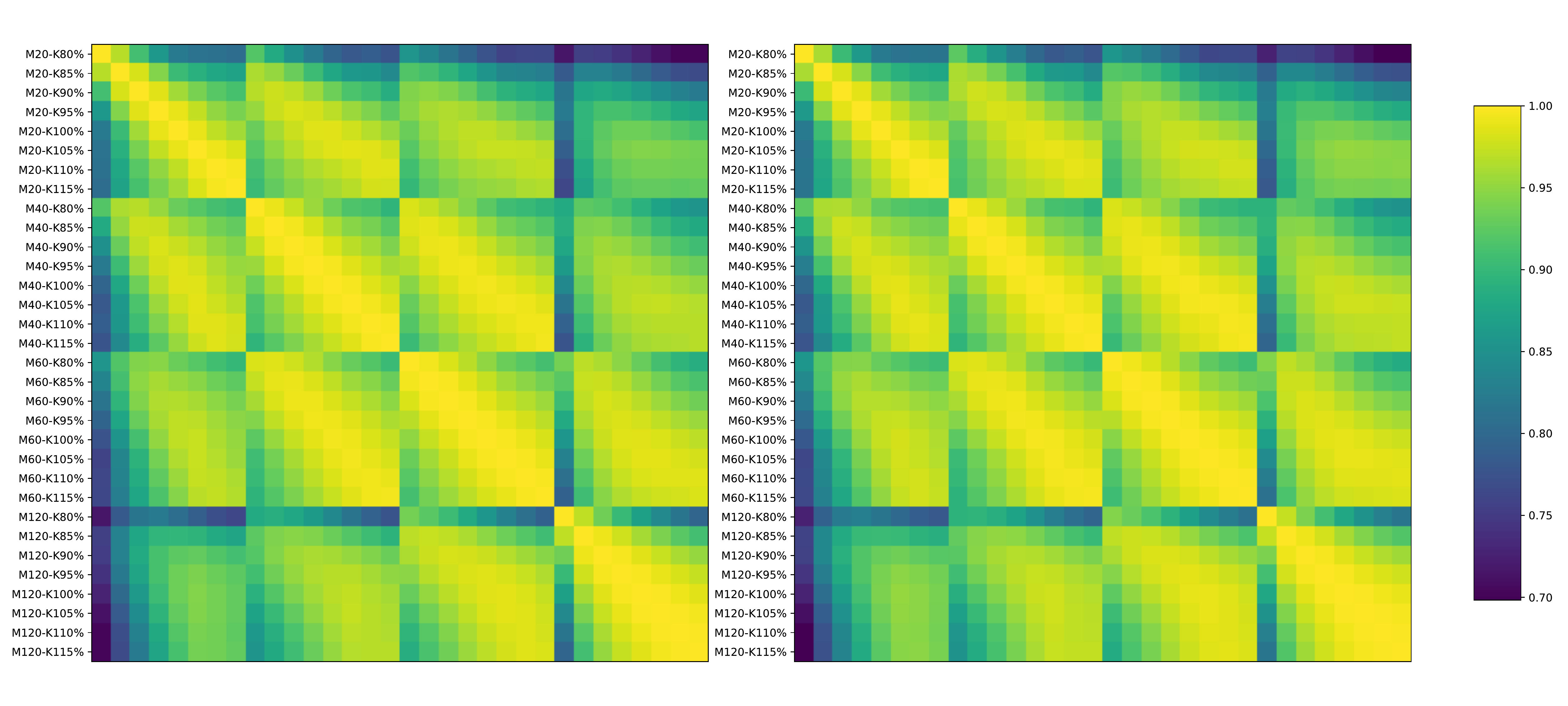}
        \caption{Cross-correlation matrix of historical (left) and generated (right) log-implied volatilities.}
       \label{fig:cross_correlation_log_ivs}
    \end{minipage}%
	\hfill
    \begin{minipage}{.45\textwidth}
        \centering
        \includegraphics[width=\textwidth]{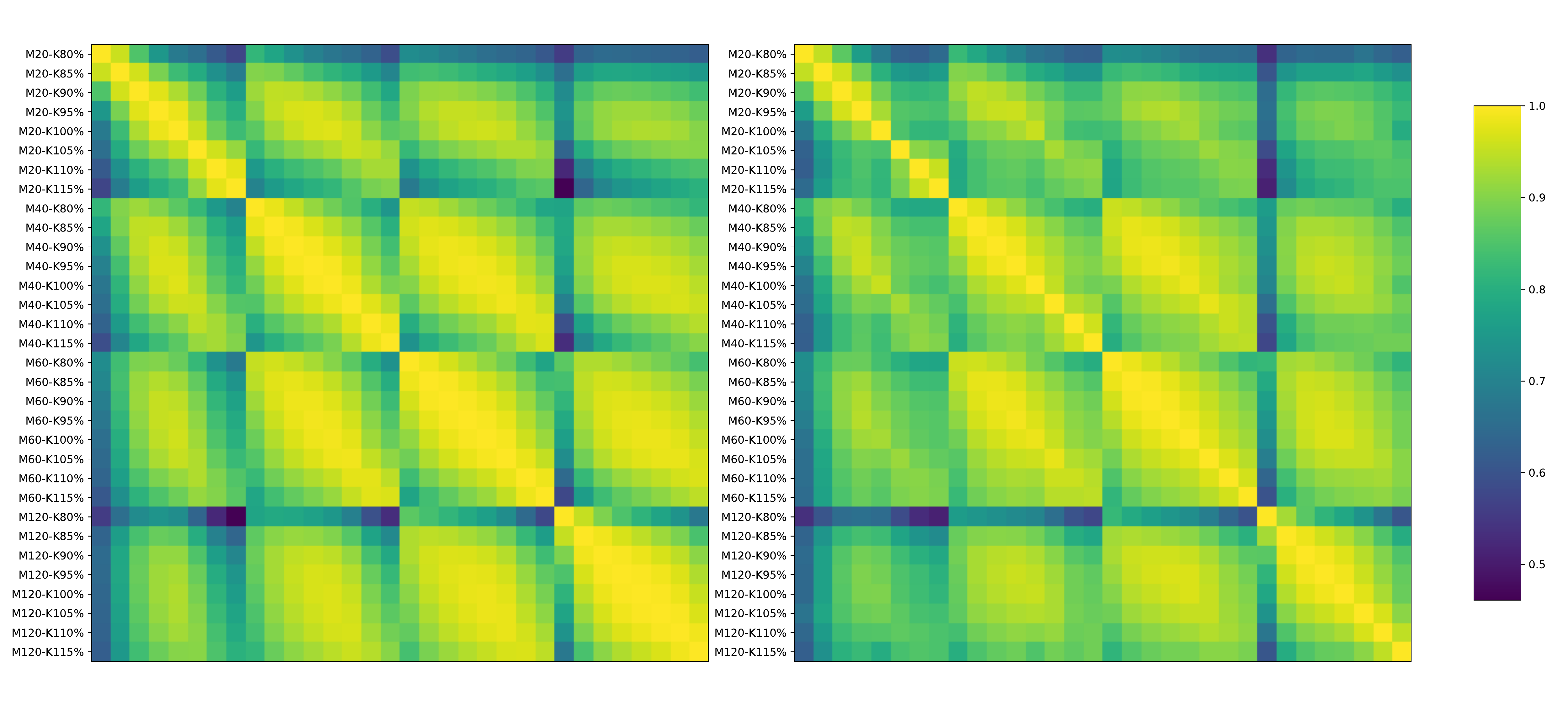}
        \caption{Cross-correlation matrix of historical (left) and generated (right) implied volatility log-returns.}
       \label{fig:cross_correlation_iv_logrtn}
    \end{minipage}
\end{figure}

\autoref{fig:cross_correlation_log_ivs} and \autoref{fig:cross_correlation_iv_logrtn} illustrate the generated and historical cross-correlation matrices for log-implied volatilities and implied volatility log-returns. In both cases the historical is approximated accurately confirming the scores in \autoref{table:full_path}.

Last viewing the dependence properties \autoref{fig:direct_gan_acf} shows that for short lags the approximation of the ACF is very good. However, for longer lags the ACF of the historical decays slower than the generated. \autoref{fig:direct_gan_acf_logrtn} displays the ACF of implied volatility log-returns and shows that the explicit GAN-trained model is able to approximate short dependencies quiet well. However, to model longer lags a larger history is necessary.

\autoref{fig:generated_implied_vols} and \autoref{fig:generated_implied_vols2} display two synthetic paths generated by the explicit GAN-trained model. Notably, the model is able to generate long-lasting periods of low volatility and periods of stress and high volatility phases. When comparing visually the synthetic paths to the historical (see \autoref{fig:historical_implied_vols}) it is difficult to discriminate them from being synthetic. 

\section{Conclusion}
In this paper, we demonstrated that the generating mechanism of implied volatilities can be closely approximated by employing adversarial training techniques. To measure the proximity of our synthetic paths to the historical we introduced a variety of scores that capture different features of implied volatilities. In \autoref{sec:numerical_results} we developed a benchmark and compared the performance of GANs against a wide range of models, training algorithms and explored the effects of compressing DLVs by using PCA.  
There we concluded that adversarial training outperforms conventional approaches such as the VAR model and qMLE training. 

Finally, our work shows for the first time that network-based models can be successfully applied to the context of generative modelling of multivariate financial time series; opening new avenues for future research and applications. 

\newpage
\bibliographystyle{plain}
\bibliography{references}

\newpage
\appendix

\renewcommand\thefigure{\thesection.\arabic{figure}} 
\section{Explicit GAN-trained model}
\setcounter{figure}{0}
\label{appendix:gan}
\subsection{Dependence properties}

\begin{figure}[!htb]
    \centering
    \begin{minipage}{.45\textwidth}
        \centering
        \includegraphics[width=\textwidth]{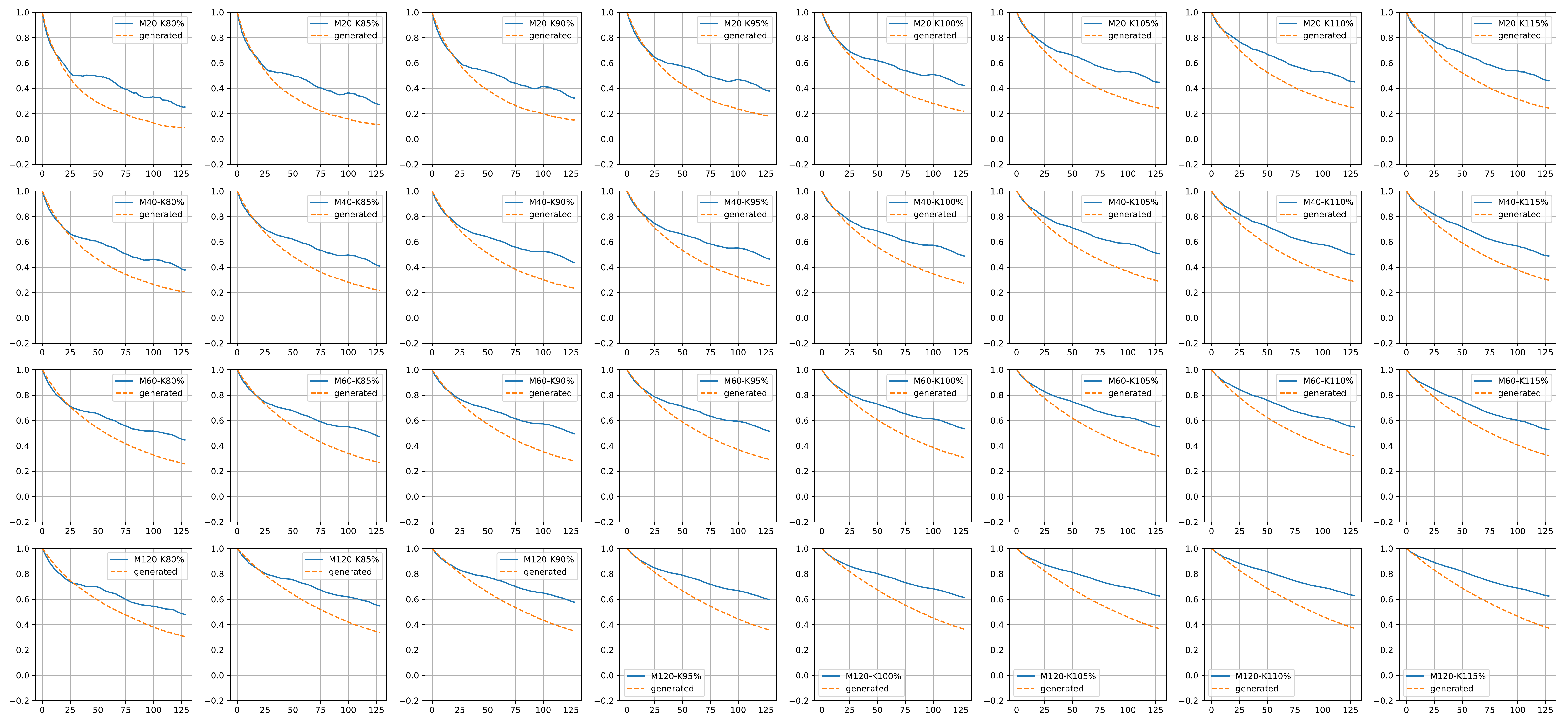}
        \caption{ACF of historical (blue) and generated (orange) log-implied volatilities.}
       \label{fig:direct_gan_acf}
    \end{minipage}%
	\hfill
    \begin{minipage}{.45\textwidth}
         \centering
        \includegraphics[width=\textwidth]{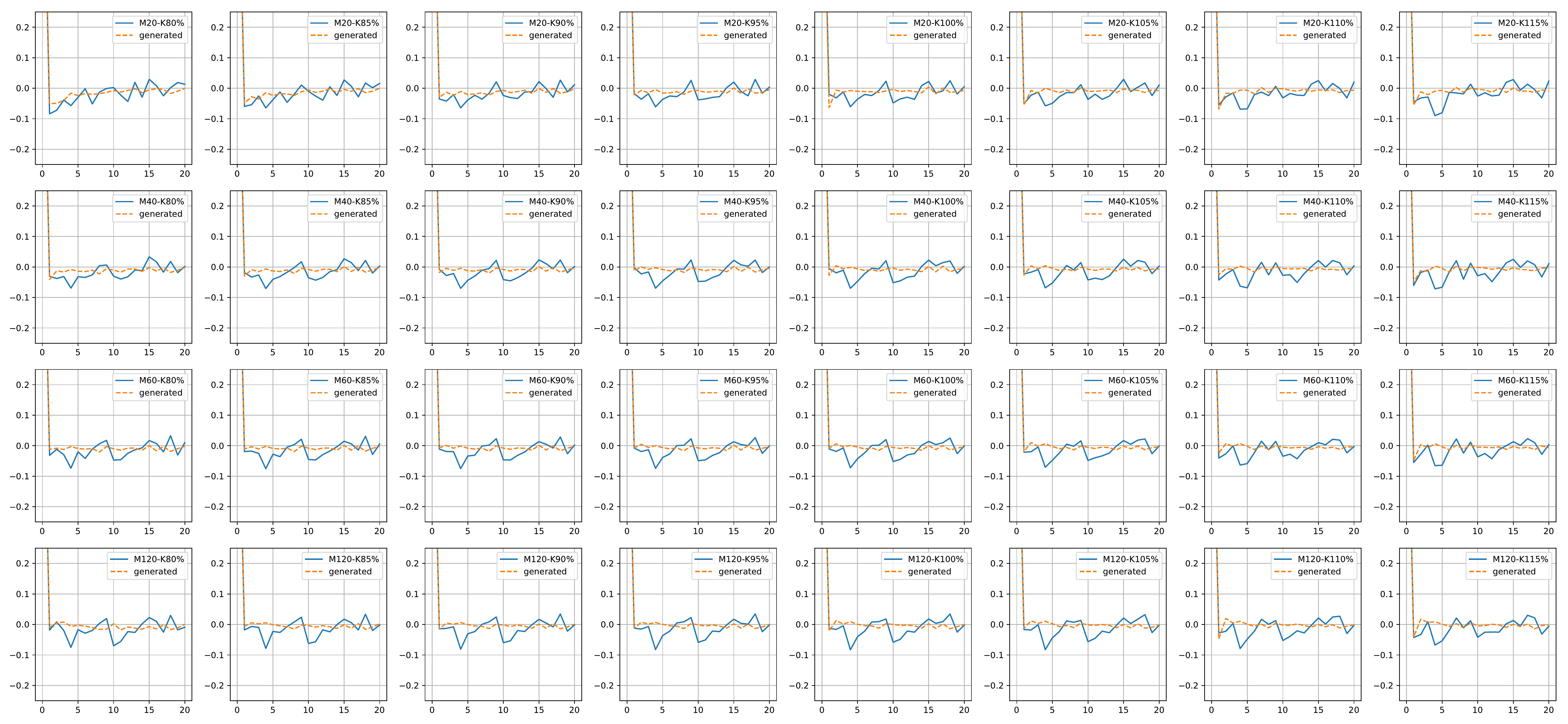}
        \caption{ACF of implied volatility log-returns of the historical (blue) and generated (orange).}
       \label{fig:direct_gan_acf_logrtn}
    \end{minipage}
\end{figure}

\subsection{Generated paths}
\begin{figure*}[!htb]
        \centering
        \includegraphics[width=\textwidth]{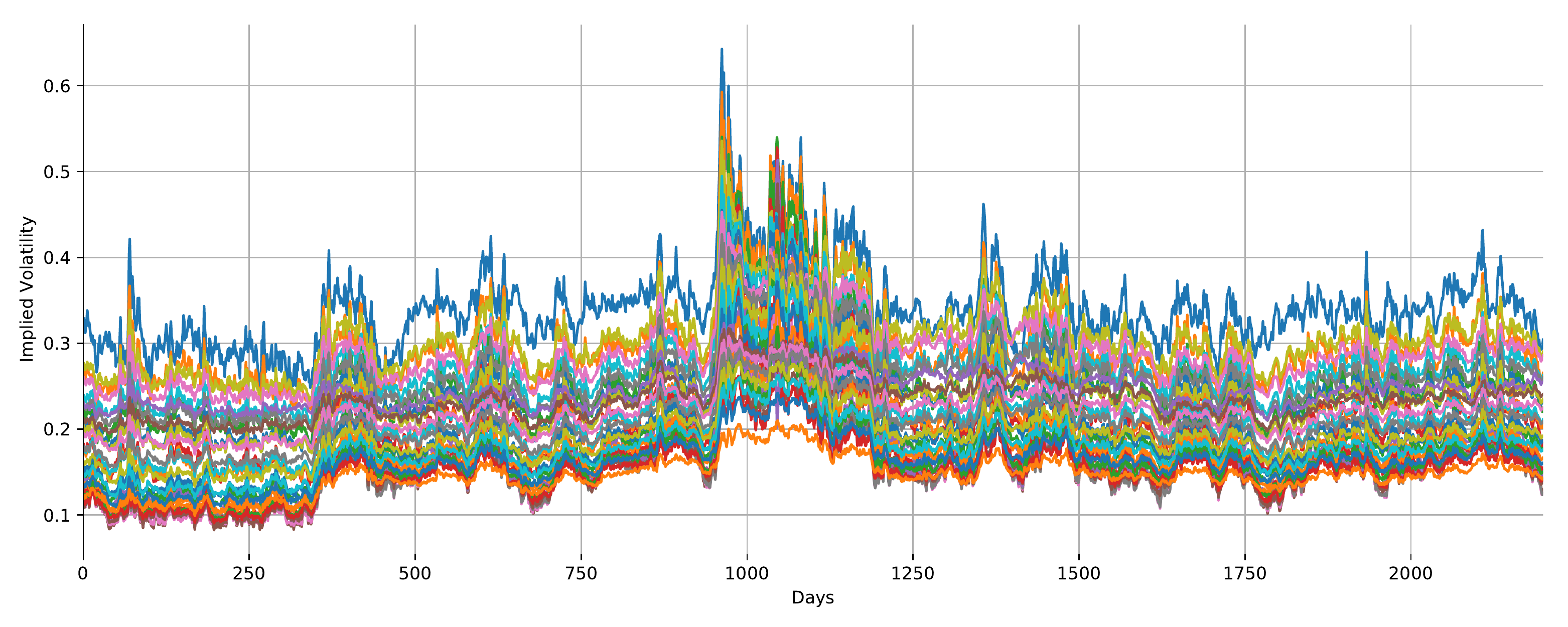}
        \caption{Explicit GAN-generated path exhibiting phases of high and low volatility.}
        \label{fig:generated_implied_vols}
\end{figure*}

\begin{figure}[!htb]
	\centering
        \includegraphics[width=\textwidth]{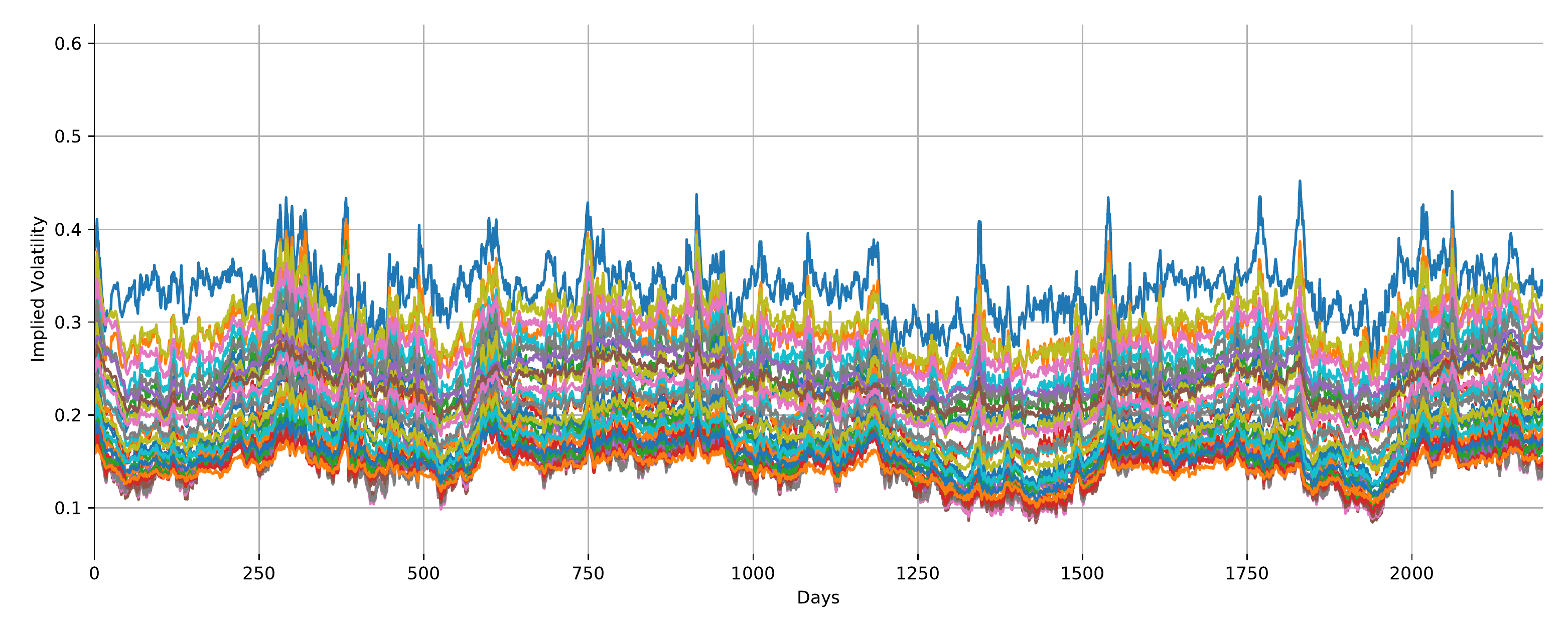}
        \caption{Explicit GAN-generated path in a low-volatility regime.}
       \label{fig:generated_implied_vols2}
\end{figure}
\newpage
\section{Discrete local volatilities}
\setcounter{figure}{0}
\label{appendix:DLVs}
DLVs define an arbitrage-free surface that is globally closest to the surface of implied volatilities. We work with DLVs instead of implied volatilities in order to satisfy arbitrage constraints. This eliminates the issue of generating option prices that contain butterfly, calendar or spread arbitrage. Having synthetic options prices that contain arbitrage is highly undesirable. An algorithm trained on these prices could potentially learn to exploit these synthetic arbitrage opportunities which do not occure in reality; yielding an unworldly algorithm that performs well on synthetic but not real data.

\begin{figure*}[htp]
        \centering
        \includegraphics[width=\textwidth]{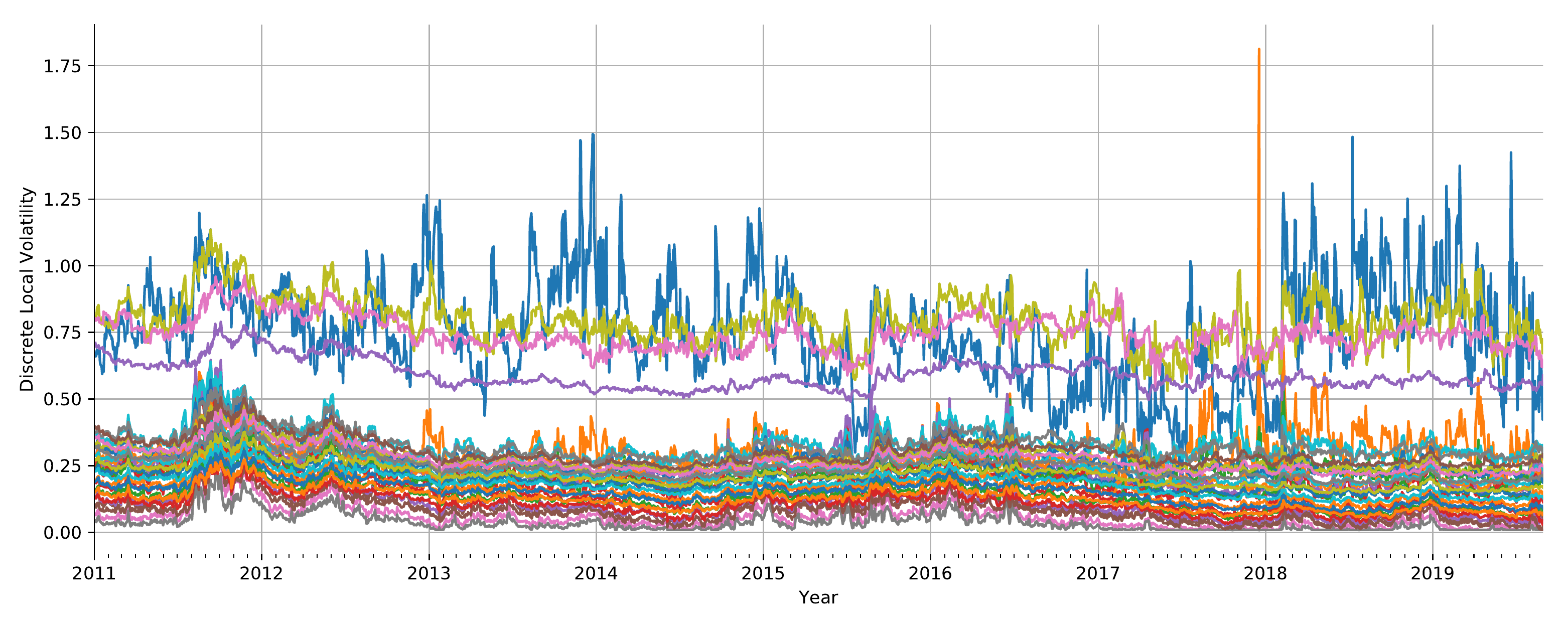}
        \caption{Discrete local volatilities of the EURO STOXX 50 for the grid $\bar \strikes \times \bar \maturities$.}
        \label{fig:historical_DLVs}
\end{figure*}

In this paper, we consider the option prices of the EURO STOXX 50 from 2011-01-03 to 2019-08-30 and for the sets of relative strikes and maturities
\begin{align*}
\bar\strikes &\coloneqq \left\lbrace 0.80, 0.85, 0.90, 0.95, 1.00, 1.05, 1.10, 1.15\right\rbrace,\\
\bar\maturities &\coloneqq \left\lbrace 20, 40, 60, 120 \right\rbrace.
\end{align*}
The historical time series of DLVs is obtained by transforming the EURO STOXX 50 implied volatilities and displayed in \autoref{fig:historical_DLVs}. 

\begin{figure}[!htb]
    \centering
    \begin{minipage}{.45\textwidth}
        \centering
        \includegraphics[width=\textwidth]{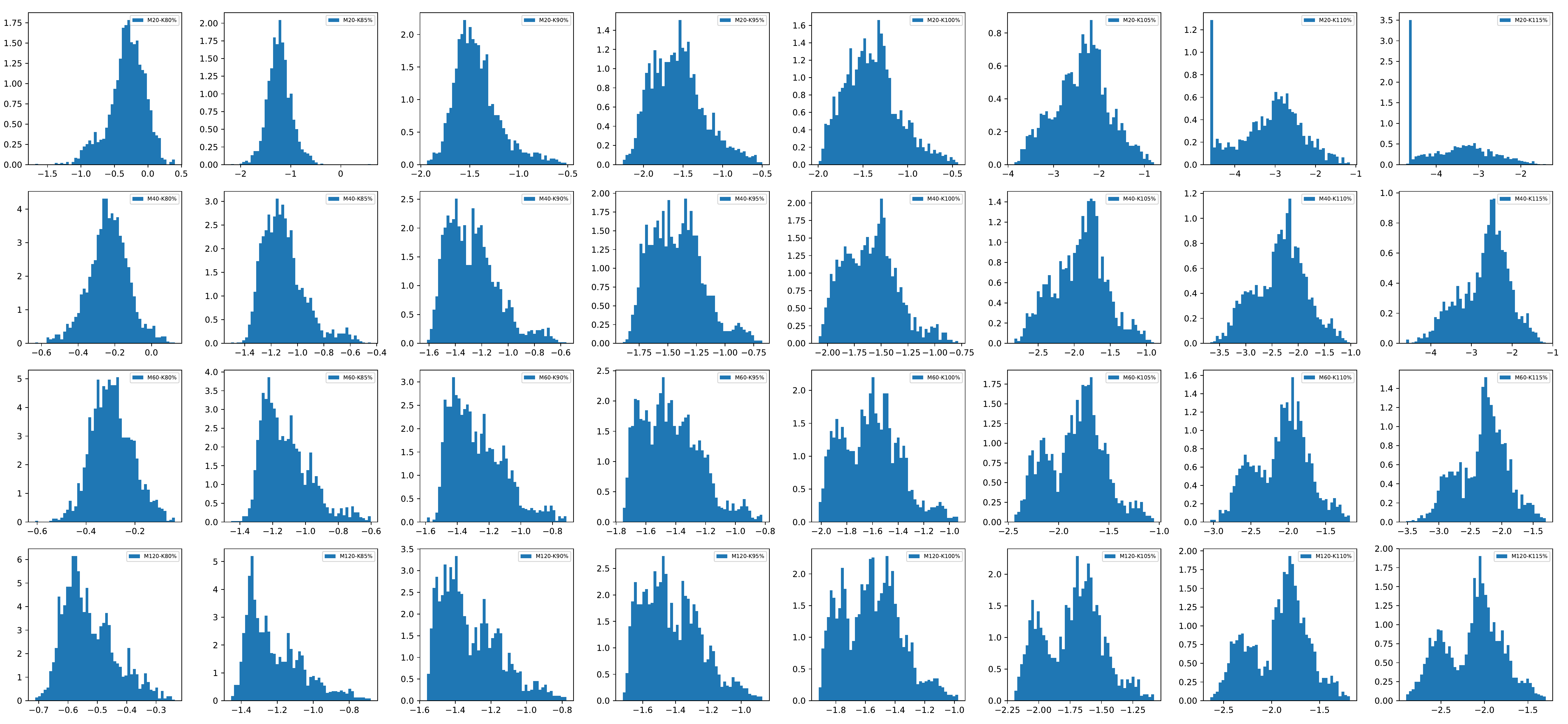}
        \caption{Empirical densities of log-DLV levels.}
        \label{fig:epdf_DLVs}
    \end{minipage}%
	\hfill
    \begin{minipage}{.45\textwidth}
        \centering
        \includegraphics[width=\textwidth]{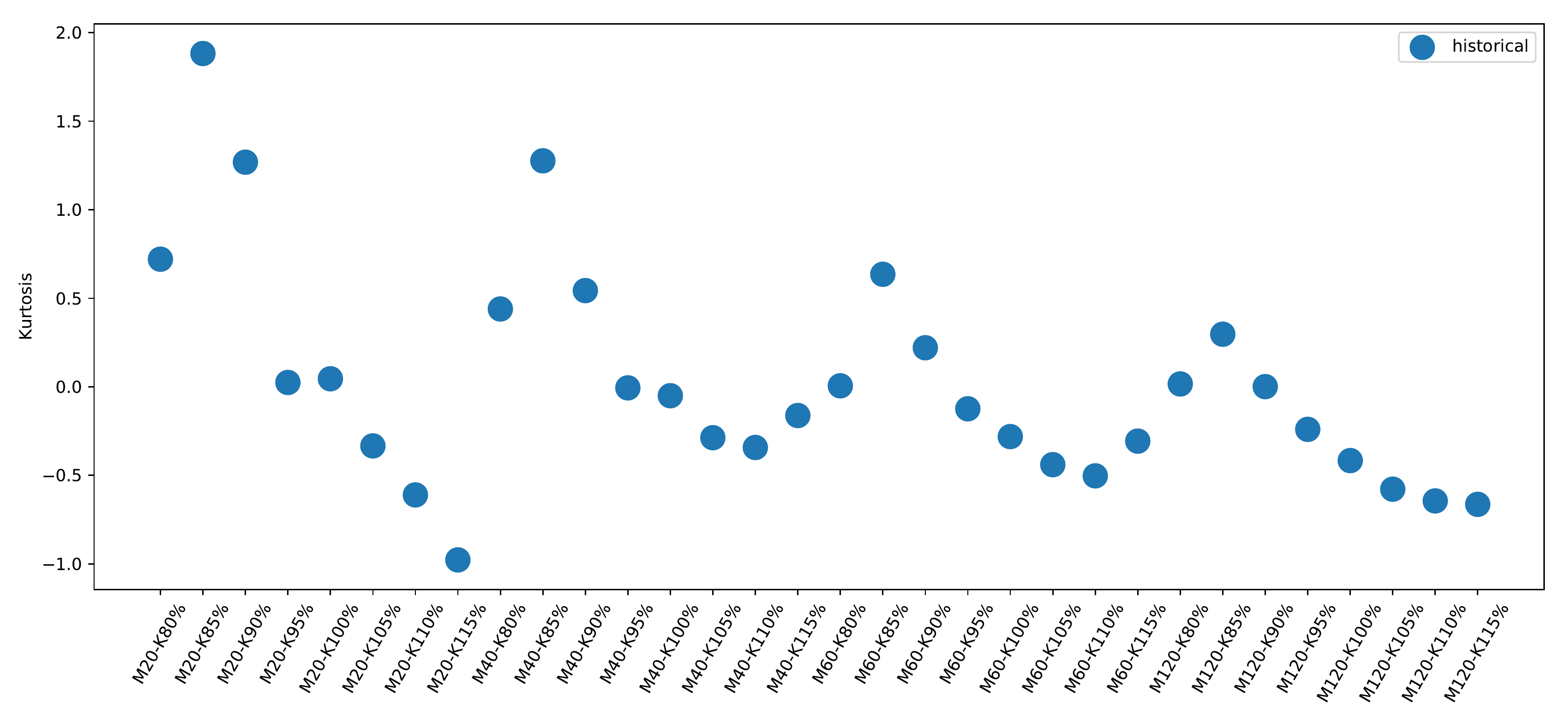}
        \caption{Kurtosis.}
        \label{fig:kurtosis_DLVs}
    \end{minipage}
\end{figure}

\autoref{fig:epdf_DLVs} shows the empirical densities of log-DLVs. Each row represents a specific maturity and columns the moneyness. Similiar to implied volatilities (see \autoref{fig:explicit_gan_distributional}) long-dated ($M \in \lbrace 60, 120 \rbrace$) out of the money (OTM) ($K \in \{105\%, 110\%, 115\%\}$) DLVs are characterised by bimodal distributions. The large buckets in the epdfs of the short-dated OTM DLVs ($M =20, \ K \in \lbrace 110\%, 115\%\rbrace$) represent the floor which is defined at $0.01$.

\begin{figure}[!htb]
 \centering
    \begin{minipage}{.45\textwidth}
        \centering
        \includegraphics[width=\textwidth]{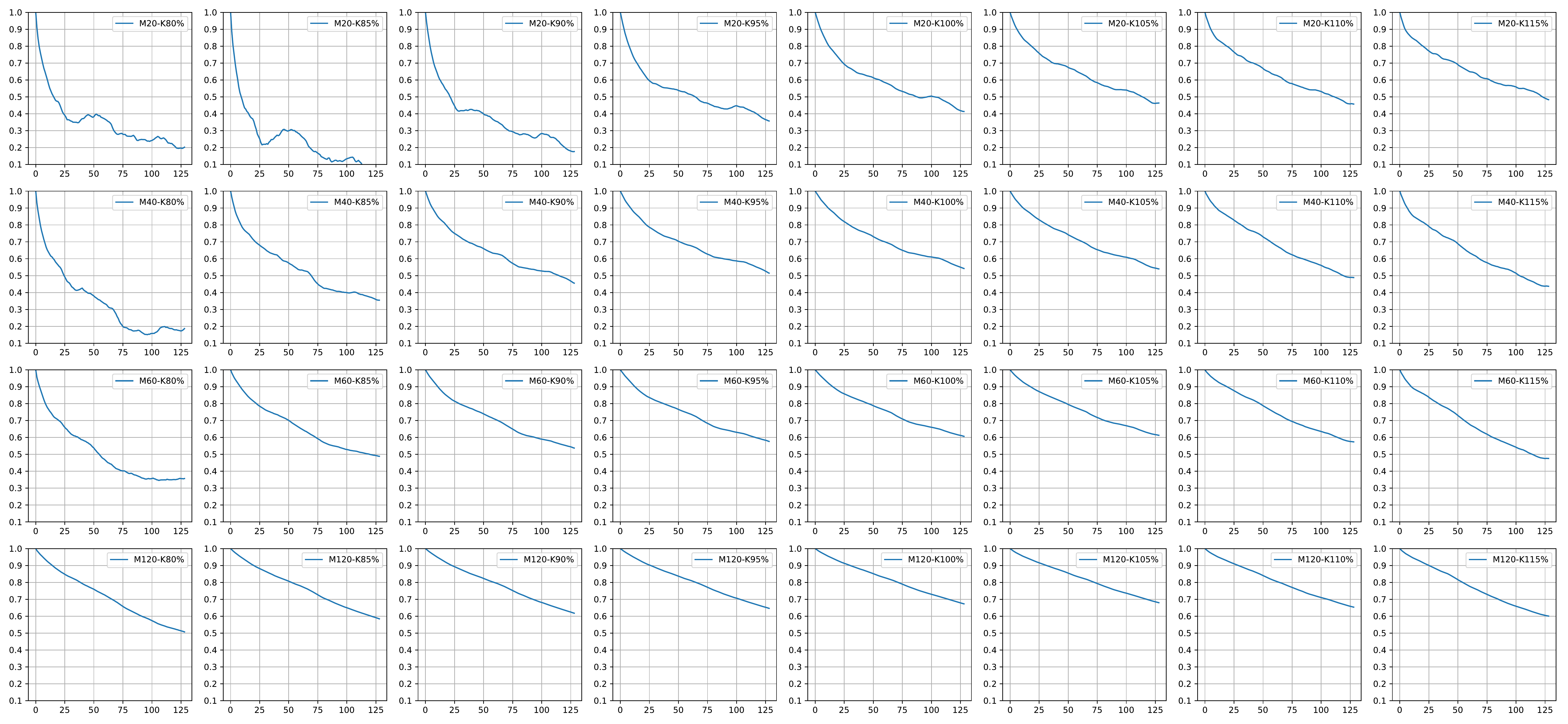}
        \caption{ACF of log-DLVs.}
        \label{fig:acf_DLVs}
    \end{minipage}%
	\hfill
    \begin{minipage}{.45\textwidth}
         \centering
        \includegraphics[width=\textwidth]{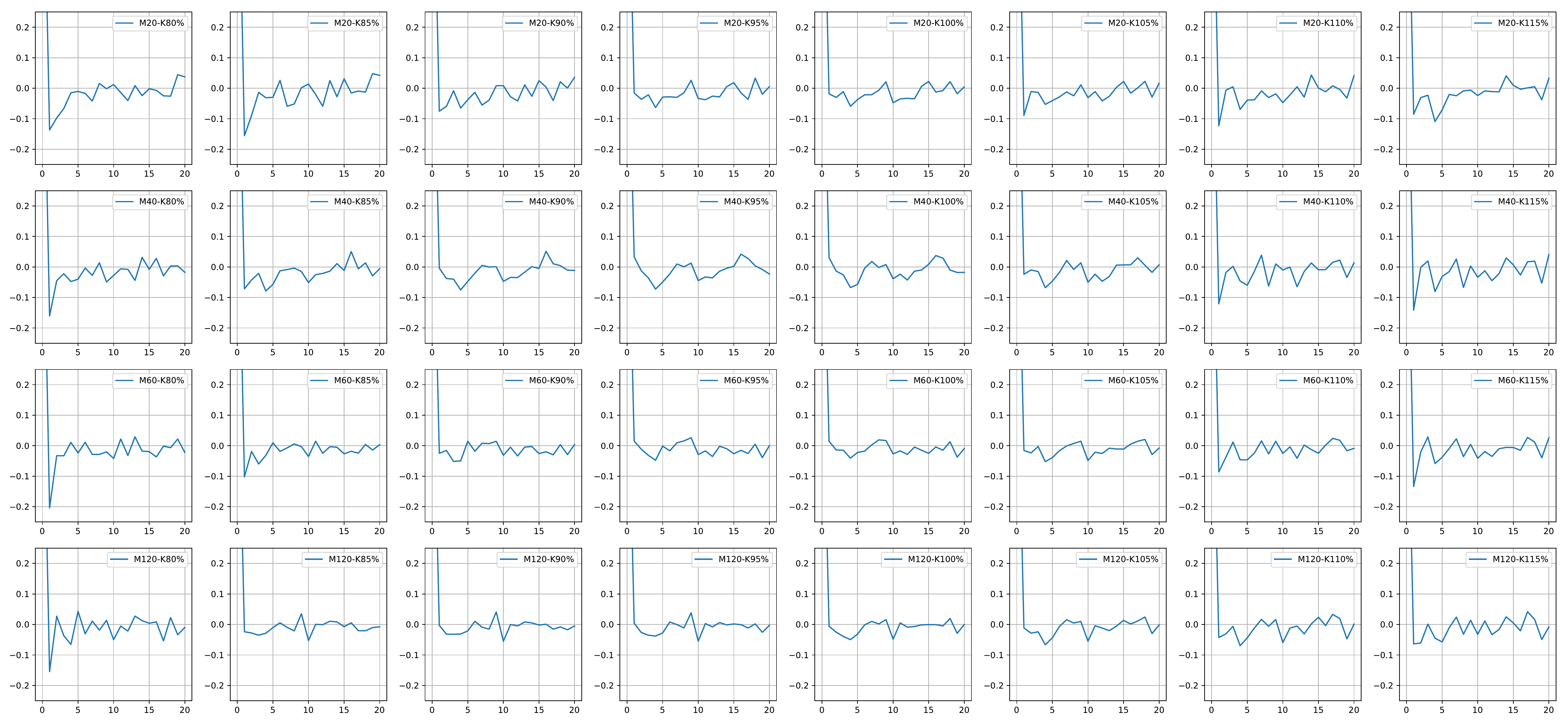}
        \caption{ACF of DLV log-returns.}
        \label{fig:acf_DLV_logrtns}
    \end{minipage}
\end{figure}

The ACFs of log-DLVs and DLV log-returns are displayed in \autoref{fig:acf_DLVs} and \autoref{fig:acf_DLV_logrtns}. Visually one can detect in \autoref{fig:acf_DLVs} that the ACFs of short-dated in the money DLVs decay faster than long-dated OTM ones.

\newpage
\section{Disclaimer}
Opinions and estimates constitute our judgement as of the date of this Material, are for informational purposes only and are subject to change without notice. It is not a research report and is not intended as such. Past performance is not indicative of future results. This Material is not the product of J.P. Morgan's Research Department and therefore, has not been prepared in accordance with legal requirements to promote the independence of research, including but not limited to, the prohibition on the dealing ahead of the dissemination of investment research. This Material is not intended as research, a recommendation, advice, offer or solicitation for the purchase or sale of any financial product or service, or to be used in any way for evaluating the merits of participating in any transaction. Please consult your own advisors regarding legal, tax, accounting or any other aspects including suitability implications for your particular circumstances.  J.P. Morgan disclaims any responsibility or liability whatsoever for the quality, accuracy or completeness of the information herein, and for any reliance on, or use of this material in any way.  Important disclosures at: \url{www.jpmorgan.com/disclosures}.

\end{document}